\pdfoutput=1
\documentclass[twocolumn]{aastex631}

\newcommand{\varv}{v}

\begin{document}
\title{Pseudo-Newtonian simulation of a thin accretion disk around a~Reissner-Nordstr\"{o}m naked singularity}

\author[0000-0002-3434-3621]{M. \v{C}emelji\'{c}}
\affiliation{College of Astronomy and Natural Sciences, SGMK Nicolaus Copernicus Superior School, Nowogrodzka 47A, 00-695, Warsaw, Poland}
\affiliation{Nicolaus Copernicus Astronomical Center of the Polish Academy of Sciences, Bartycka 18, 00-716 Warsaw, Poland}
\affiliation{Research Center for Computational Physics and Data Processing, Institute of Physics, Silesian University in Opava, Bezru\v{c}ovo n\'am.\ 13,\\ CZ-746-01 Opava, Czech Republic}
\affiliation{Academia Sinica, Institute of Astronomy and Astrophysics, P.O. Box 23-141,
Taipei 106, Taiwan}

\author{W. Klu\'{z}niak}
\affiliation{Nicolaus Copernicus Astronomical Center of the Polish Academy of Sciences, Bartycka 18, 00-716 Warsaw, Poland}

\author[0000-0002-9475-5193]{R. Mishra}
\affiliation{College of Astronomy and Natural Sciences, SGMK Nicolaus Copernicus Superior School, Nowogrodzka 47A, 00-695, Warsaw, Poland}
\affiliation{Nicolaus Copernicus Astronomical Center of the Polish Academy of Sciences, Bartycka 18, 00-716 Warsaw, Poland}

\author[0000-0002-8635-4242]{M. Wielgus}
\affiliation{Instituto de Astrof\'{i}sica de Andaluc\'{i}a-CSIC, Glorieta de la Astronom\'{i}a s/n, E-18008 Granada, Spain}

  \begin{abstract}
  We present the first numerical simulations of a thin accretion disk around a Reissner-Nordstr\"{o}m (RN) naked singularity (a charged point mass). The gravity of the RN naked singularity is modeled with a pseudo-Newtonian potential that reproduces exactly the radial dependence of the RN Keplerian orbital frequency; in particular, orbital angular velocity vanishes at the zero gravity radius and has a maximum at 4/3 of that radius. Angular momentum is transported outwards by viscous stresses only outside the location of this maximum. Nonetheless, even at that radius, accretion proceeds at higher latitudes, the disk having thickened there owing to excess pressure.
  The accretion stops at a certain distance away from the singularity, 
  with the material accumulating in a toroidal structure close to the zero-gravity sphere. The shape of the structure obtained in our simulations is reminiscent of fluid figures of equilibrium analytically derived in full general relativity for the RN singularity. The presence of a rotating ring, such as the one found in our simulations, could be an observational signature of a naked singularity. For charge to mass ratios close to but larger than unity, the inner edge of the quasi-toroidal inner accretion structure would be located well within the Schwarzschild marginally stable orbit (ISCO), and the maximum orbital frequency in thin accretion disks would be much higher than the Schwarzschild ISCO frequency.
  \end{abstract}


%
\section{Introduction}

The existence of naked singularities (NkS) is a subject of continuing theoretical debate. They appear as a solution to vacuum field equations in general relativity (GR) as the superspinar Kerr solution (for a compact object endowed with mass and spin alone)    or the highly charged Reissner-Nordstr\"{o}m solution (for a compact object endowed with mass and charge alone). NkS also occur for high charges and/or high spins in the Kerr-Newman metric (compact object endowed with mass, spin, and charge). Naked singularities appear also in many non-GR theories of gravity such as the $f(R)$ gravity model \citep{Buch70, Starob80}, and in non-vacuum spacetimes \citep[and references therein]{Nojiri17}. The cosmic censorship conjecture, stating that NkS cannot be formed in nature, i.e., that gravitational collapse to a singularity will always be accompanied by the formation of an event horizon \citep{Penrose69}, is hard to formulate precisely, as there already exist models of gravitational collapse leading to NkS for certain initial conditions \citep{Shapiro1991,Woosley93,Joshi2014}. It is unclear if ``realistic" initial conditions could lead to a collapse to a NkS \citep{MacFad99}; however, it is also debatable what constitutes ``sufficiently realistic'' initial conditions. For instance, it is now apparent that the Oppenheimer-Snyder \citep{Openh39} black hole (BH) collapse solution does not meet the initial conditions requirements that would follow from realistic stellar evolution calculations, but there is no doubt that it is fundamentally correct. 

The uncertain ontological status of NkS encourages studies of detectability of such objects. A particularly topical question is whether their properties could allow to make a robust distinction between a BH and NkS of the same mass. It has been shown that the lensing properties of NkS differ from those of black holes \citep[and references therein]{Virbhadra1998,VirbhadraEllis2002,Wagner2023,Wagner2024,Vagnozzi23}. Event-horizon scale imaging of supermassive black holes M87* and Sagittarius~A* (Sgr~A*) by the Event Horizon Telescope (EHT) invigorated this debate recently, with an increasing number of papers discussing the appearance of accreting systems around naked singularities \citep{Pugliese11, VSK14, 2015RN, Gyulchev2020, VK23, MishraK23, KK24, Dihingia2024}. On the other hand, the accretion onto a naked singularity itself is a theoretical problem that may require going beyond the classic gravity theories in order to describe the physics in the immediate vicinity of the singularity.
\begin{figure*}[ht!]
\centering
\includegraphics[width=0.99\columnwidth]{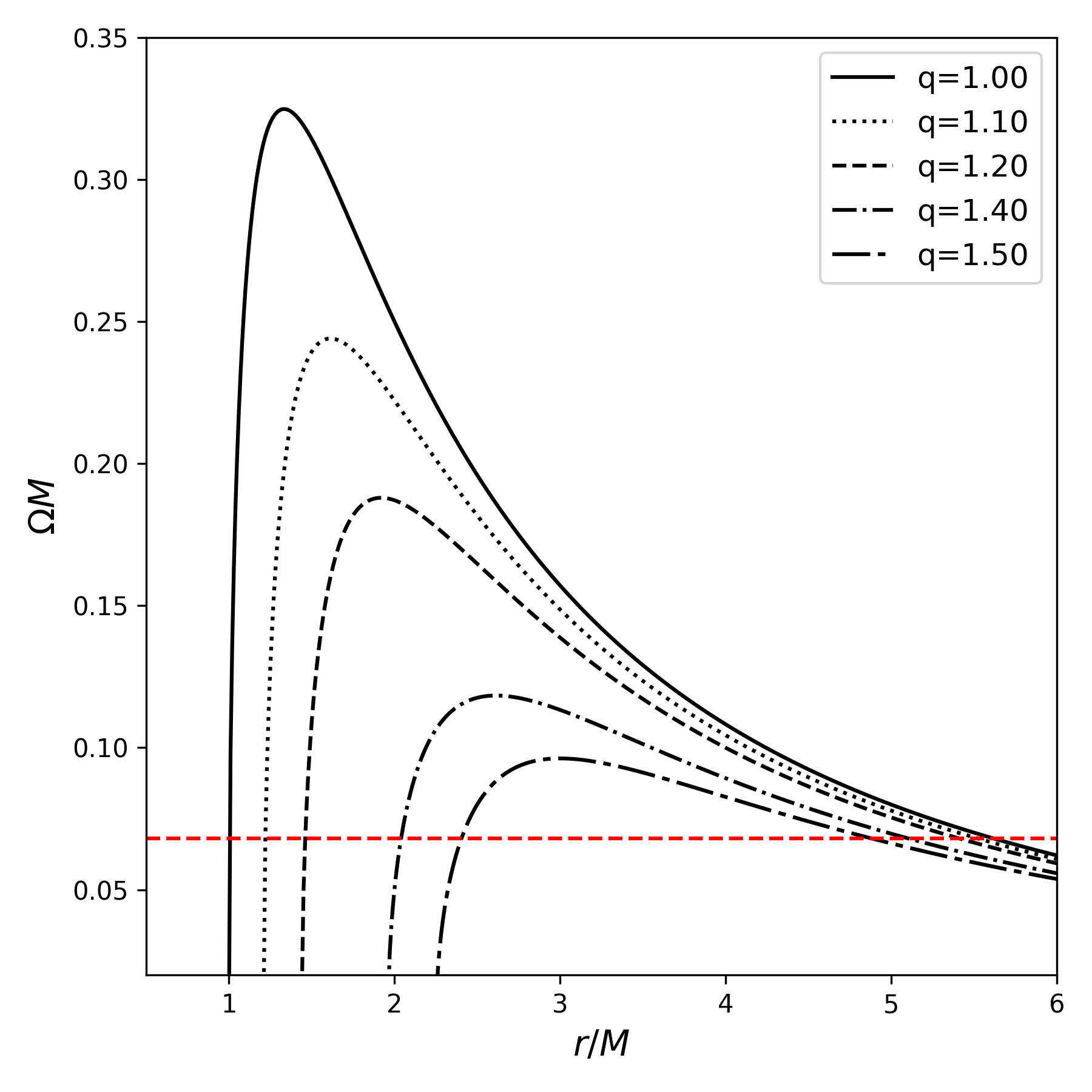}
\includegraphics[width=0.99\columnwidth]{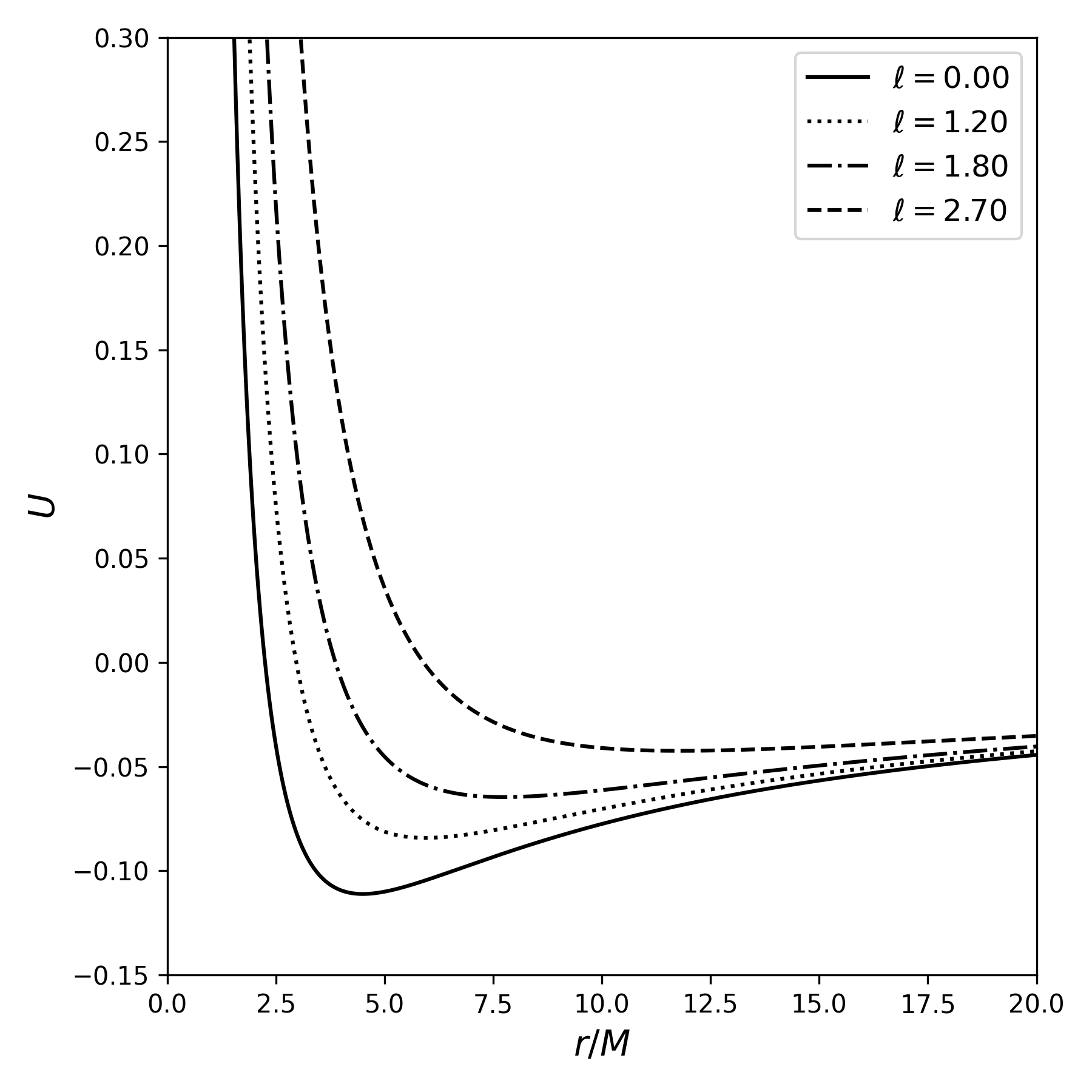}
\caption{{\it Left panel}: square of the RN test-particle orbital frequency as a function of $r/M$ (Eq.~\ref{omgrn}) for the charge parameter values $q = 1.00, 1.10, 1.20, 1.40, 1.50$, arranged from top to bottom. The red dashed line represents the Schwarzschild value of the  ISCO angular frequency. {\it Right panel}: the pseudo-Newtonian effective potential $U$ from Eq.~\ref{effect} with $q=1.5$ for different values of the specific angular momentum $\ell$.
}
\label{pl1}
\end{figure*}
The EHT results show unequivocally that at the center of our Galaxy, as well as the nearby giant elliptical galaxy M87, sits a supermassive compact object \citep{M87P1,SgraP1}, quite likely a black hole \citep{M87P6,SgrAP6}. EHT images reveal ring-like structure at the center of the Milky Way, in the Sgr~A* object, with an extent of only a few gravitational radii for the independently and precisely determined mass to distance ratio of the central object \citep{Gravity2022}. A similar compact feature was observed in M87*. The ring is interpreted as the lensed image of the innermost part of the accretion flow \citep{M87P5,SgrAP5}, including a light ring surrounding a black hole, related to the black hole ``shadow'', or silhouette \citep[e.g.,][]{Wielgus2021,Paugnat2022}. In addition to the Kerr metric, the images allow for interpretation with more exotic spacetimes \citep{Vincent2021}.
Another suggestive, but as yet unverified, interpretation of the image proposed recently  \citep{MishraRNakSing24} is that it might correspond to a ring-like fluid equilibrium structure around a NkS.
Crucially, the observations now allow quantitative tests of various spacetime metrics (or theories of gravity) theoretically allowed for supermassive compact objects, through comparison of observations in the electromagnetic domain with the predicted properties of hot matter orbiting compact sources of gravity. This motivates further studies of fluid motion and accretion for spacetimes more general than the familiar Schwarzschild and Kerr solutions of GR.

In this work we study a well-known GR solution, the Reissner-Nordstr\"{o}m (RN) spacetime, which describes the gravity around a spherically symmetric, electrically charged object. The configurations of (hydrostatic) equilibrium for a~fluid orbiting a RN naked singularity have already been determined for uniform angular momentum distributions in \cite{MishraK23} and \cite{MishraKK24}. Here, we attempt to determine the outcome of thin-disk accretion onto a RN singularity through numerical simulations. Generally, when fluid accretes onto a central object, the angular momentum distribution is seldom uniform, and in fact the angular momentum is typically transported outwards by viscous torques.

To allow for rapid and computationally inexpensive simulations we introduce a pseudo-Newtonian potential for the 
RN naked singularity, capturing the relevant properties of the metric. We use this potential in the hydrodynamic regime with the well-tested and widely used PLUTO code \citep{m07} to evolve in time a thin accretion disk that is initially placed at a large distance from the singularity. To our knowledge, this is the first simulation of a thin accretion disk around a naked singularity. This set-up is suitable for investigating moderate mass-accretion rate systems involving thin disk accretion onto a compact object. We note that in recent GR simulations of accretion onto RN and Kerr NkS \citep[][respectively]{KK24,Dihingia2024} the accretion structure was a thick torus with a cusp, and not a thin disk. Further, the \cite{KK24} RN study was performed for a value of the charge parameter very close to the value that allows a black hole solution, to take advantage of the presence of unstable circular orbits (and the ISCO), whereas here we study a thin disk for a generic RN NkS where no unstable circular orbits (and no photon orbits) are present, i.e., for values of the charge parameter satisfying $q \equiv Q/M >\sqrt{9/8}$ \citep{MishraKK24}.

In Section~\ref{RNnaks} we briefly discuss the pseudo-potential for RN naked singularity, and present our results in Section~\ref{rezs2}, with the numerical setup detailed in Section~\ref{stp}. We conclude in Section~\ref{conc}.

\section{Pseudo-Newtonian potential for RN spacetime}\label{RNnaks}
\leavevmode
\begin{figure*}
\centering
\includegraphics[width=0.99\columnwidth]{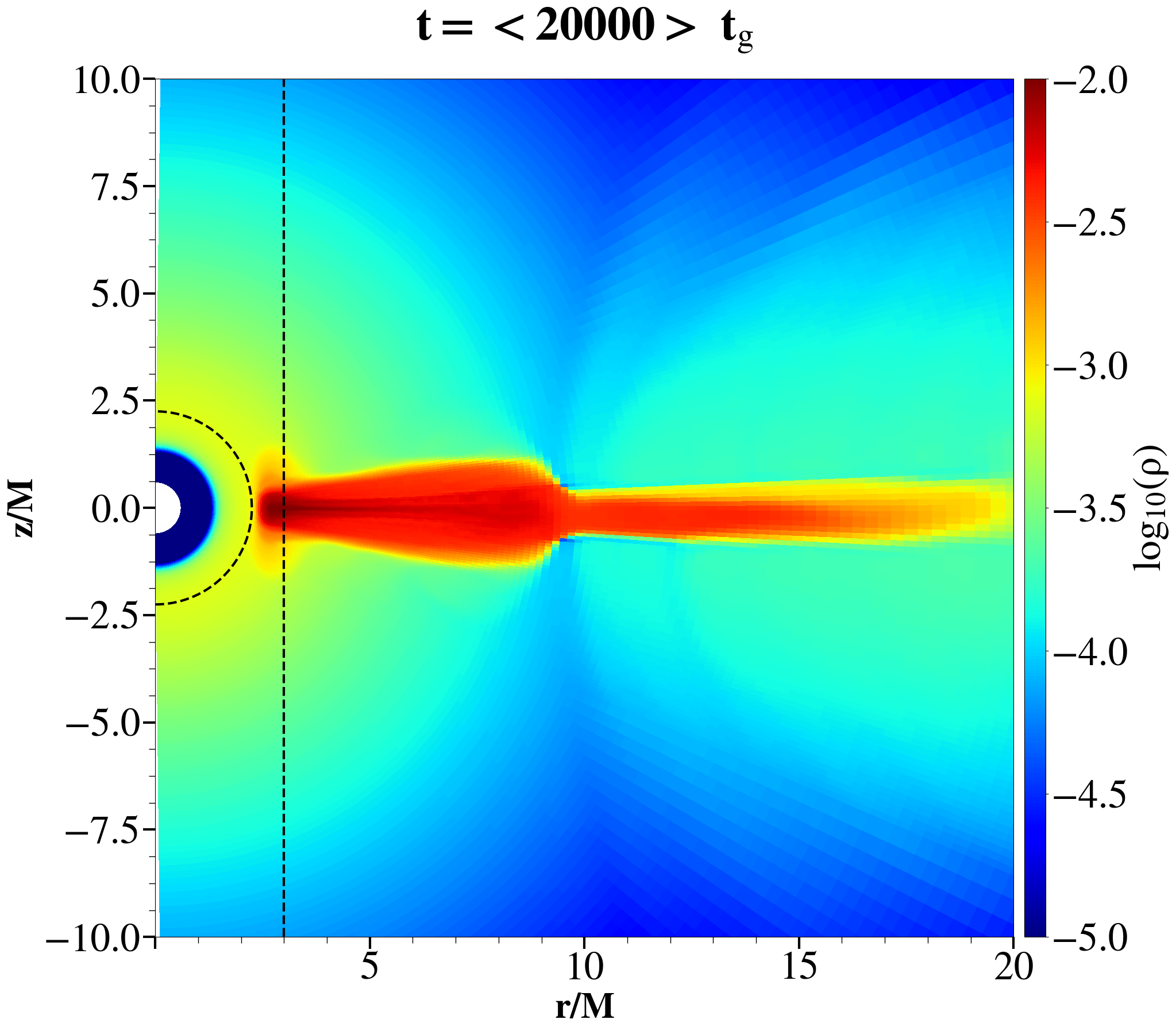}
\includegraphics[width=0.85\columnwidth]{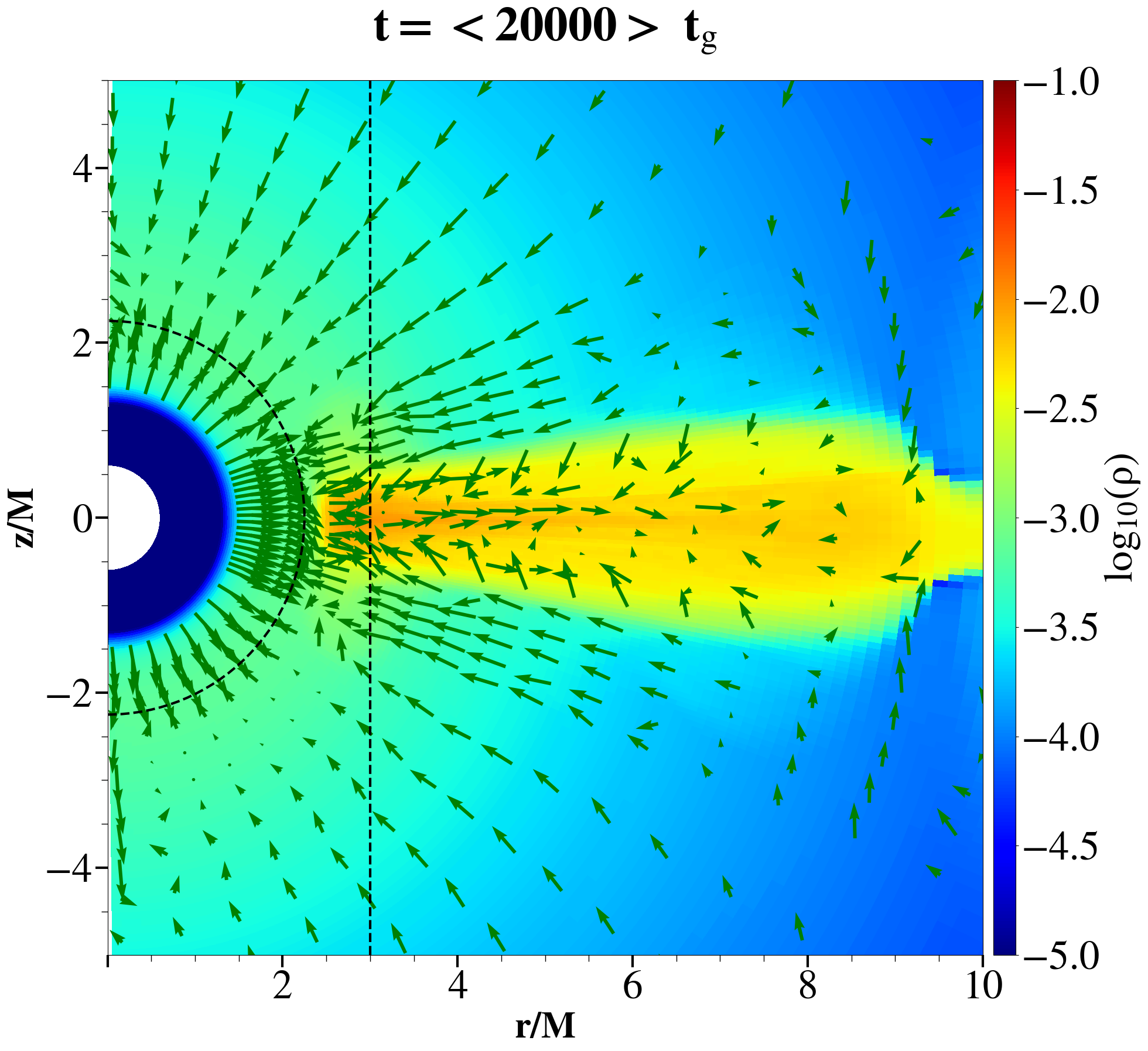}
\caption{Gas density in the simulation with $q=1.5$, obtained as an average over the time interval of $t\in [19000, 21000]\ t_{\rm g}$ centered at $t=<20000>~t_{\rm g}$, where $t_{\rm g} = r_{\rm g}/c$. {\it Left panel}: result at $t=<20000>~t_{\rm g}$, with the zero-gravity radius $r_0$ marked with the dashed half-circle and the radius of $4r_0/3$, at which the Keplerian angular velocity $\Omega$ attains a maximum, marked with the straight black dashed line. {\it Right panel}: a zoom into the inner region of the accretion flow from the left panel, with the poloidal gas velocity vectors indicated by green arrows.}
\label{rho1c5}
\end{figure*}

The RN metric describes the gravity of an electrically charged spherically symmetric body in general relativity \citep{Reiss16,Nords18}. We use gravitational units ($G=c=1$), in which the gravitational radius, $r_\mathrm{g}=GM/c^2$, and the gravitational time $t_\mathrm{g}=GM/c^3$, become simply $r_\mathrm{g} = t_\mathrm{g} = M$. The square of the line element $ds$ is
\begin{eqnarray}
ds^2=-f(r)\,dt^2+\frac{1}{f(r)}dr^2+r^2\left(d\theta^2+\sin^2\theta\, d\phi^2\right)\quad \label{rnline1}\\
{\rm with}\quad f(r)=1-\frac{2M}{r}+\frac{Q^2}{r^2}\equiv 1-2\frac{M}{r}+q^2\left(\frac{M}{r}\right)^2 
\label{rnline2}
\end{eqnarray}
where $M$ is the mass and $Q$ the electric charge of the gravitating body in gravitational units, and $q = Q/M$ is the dimensionless charge parameter.
For $|q|>1$ the spacetime is that of a (point-like, and hence spherically symmetric) naked singularity, i.e., in principle all space at $r>0$ is accessible to observations. For $|q| >\sqrt{9/8}$ test-particle circular orbits exist for the range of radii from $r=\infty$ all the way down to 
\begin{equation}
r_0=Q^2/M = q^2 M, 
\end{equation}
the zero-gravity radius, where the test particle can remain at rest \citep[for lower radii gravity is repulsive;][]{VK23}. The Keplerian (circular timelike geodesic) orbital angular frequency of a test-particle in this regime is given by 
\begin{equation}
\Omega_\mathrm{RN}(r) = \frac{u^\phi_{\rm Kep}}{u^t_{\rm Kep}} =\sqrt{\left(1-\frac{r_0}{r}\right)\frac{M}{r^3}}
\label{omgrn}
\end{equation}
and attains a maximum at $r_\mathrm{\Omega max}=4r_0/3$ \citep{Pugliese11,MishraRNakSing24}.

In the simulations reported here, we use a simple ("pseudo-Newtonian") potential 
\begin{equation}
V(r)=-\frac{M}{r}+\frac{Q^2}{2r^2}
\label{rnline2b}
\end{equation}
that reproduces in Newtonian mechanics the location of the zero-gravity radius as a minimum of $V(r)$, that is $V'(r_0) = 0$. The effective potential for a test particle with a specific angular momentum $\ell$ can be then calculated as 
\begin{equation}
U(r) = V(r)+\frac{\ell^2}{2r^2}=-\frac{M}{r}+\frac{Q^2}{2r^2} +\frac{\ell^2}{2r^2}. 
\label{effect}
\end{equation}
The effective potential given by Eq.~\ref{effect} reproduces in the Newtonian formalism the \emph{exact} functional form of the Keplerian orbital frequency, Eq.~\ref{omgrn}, in the RN spacetime, $\Omega\equiv\Omega_\mathrm{RN}$ for all $r>r_0$, including the maximum at $r=4r_0/3$ and zero value at $r_0$. To see that, it is enough to solve $U'(r) = 0$ for $\ell$ and use the Newtonian relation $\ell(r)=r^2\Omega(r)$. Examples of the RN Keplerian angular frequency as a function of the orbital radius for various values of the dimensionless parameter $q$ are shown in Fig.~\ref{pl1} (left panel) together with the effective (pseudo-Newtonian) potential $U(r)$ for various values of the specific angular momentum $\ell$ (right panel).

It is important to note that with $Q=0$ we recover the standard Newtonian potential, rather than, for example, Paczy\'nski-Wiita pseudo-potential model of the Schwarzschild spacetime \citep{PW80}. This is because Paczynski-Wiita pseudo-potential reproduces exactly the radial dependence of the Keplerian orbital {\it angular momentum} in the Schwarzschild spacetime, as well as the location of the innermost stable circular orbit and the marginally bound orbit, but not the Keplerian orbital velocity \citep{Abramowicz09}, especially not in the innermost orbits. In this work, we have a somewhat different aim of giving an effective model of accretion on the RN naked singularity by reproducing observables, such as the zero gravity sphere location in terms of the circumferential radius, and radial dependence of the Keplerian {\it angular  velocity}. We recall that in the regime of interest here ($|q| >\sqrt{9/8}$) neither the ISCO nor the marginally bound orbit exist.
\begin{figure*}[ht!]
\centering
\includegraphics[width=\columnwidth]{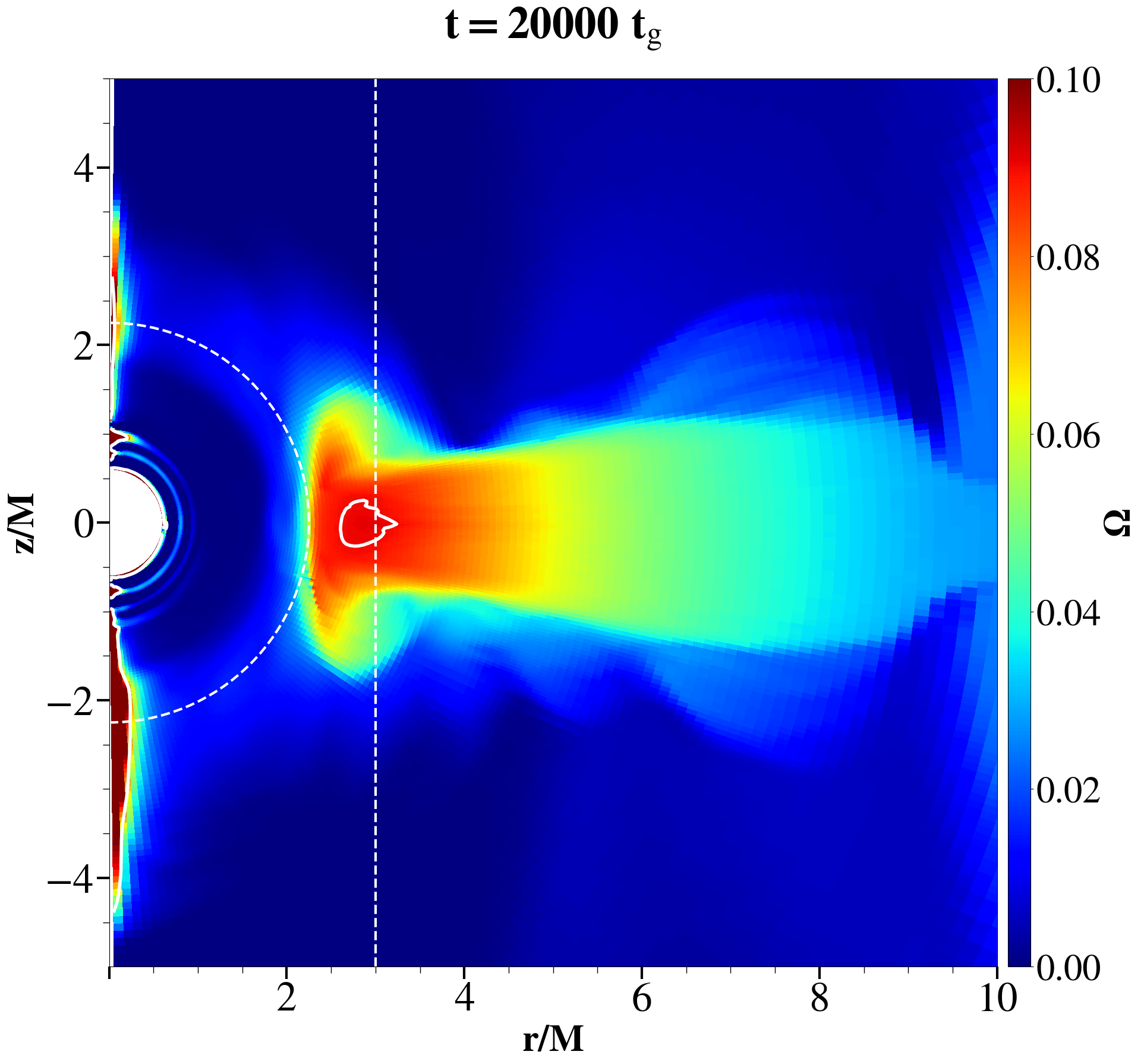}
\includegraphics[width=0.9\columnwidth]{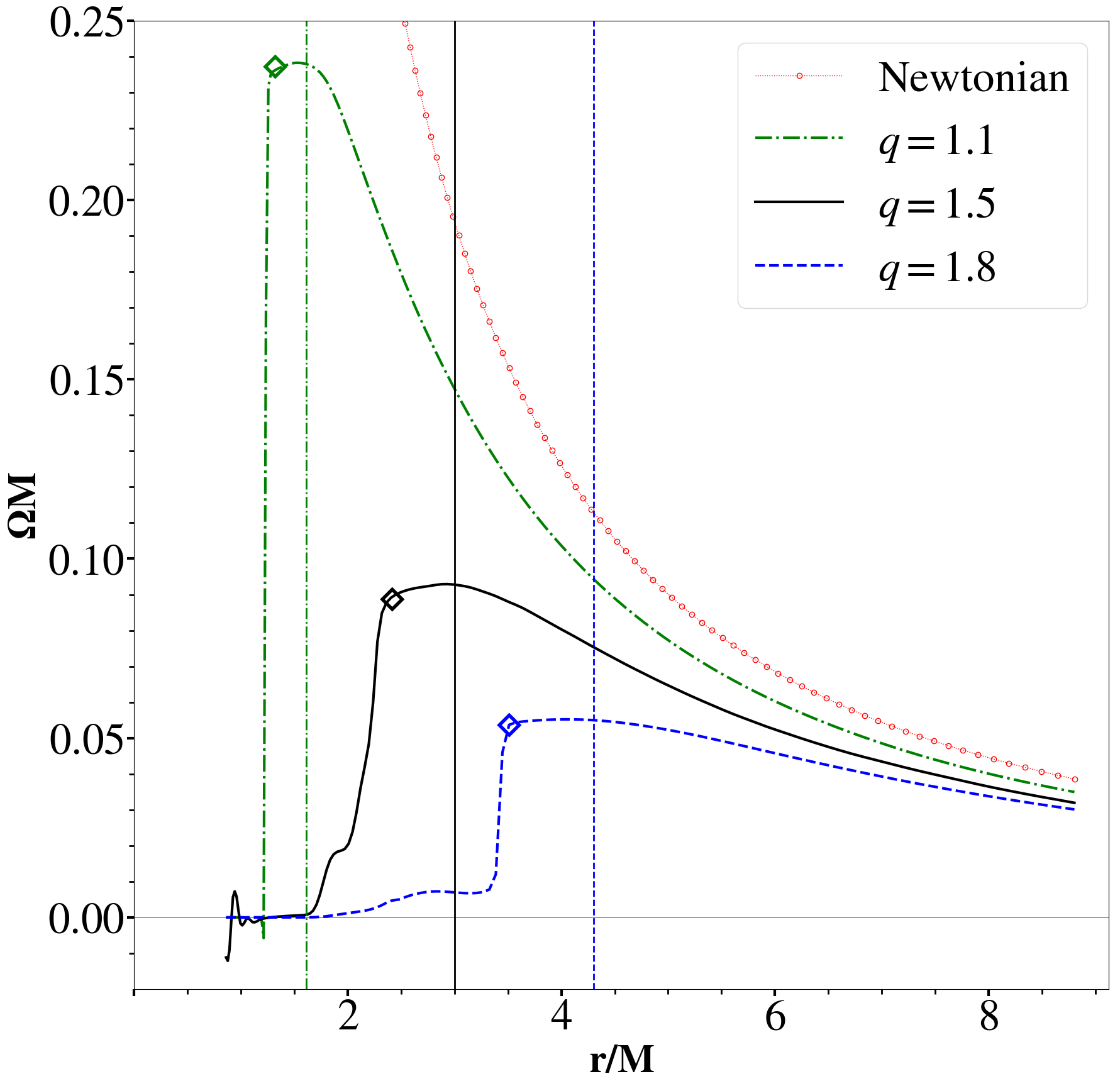}
\caption{
{\it Left panel:} the angular velocity 
$\Omega=\varv_\phi /(r\cos{\theta})$, in a linear color grading in a snapshot at $t=20000\,t_{\rm g}$ in our simulation with $q=1.5$. The contour of $\Omega=0.09/M$, within which is located the test-particle orbital frequency value of $\Omega_{\rm max}$ at $r/M=4q^2/3=3$, is shown with the white solid curve. The white dashed circular line indicates the zero-gravity sphere. {\it Right panel:} 
$\Omega(r)$ in the equatorial plane
for the RN metric with $q=1.1$, 1.5 and 1.8, in dot-dashed (green), solid (black) and dashed (blue) lines, respectively. The dotted (red) line follows the Newtonian profile of $\Omega M$ for stable test particle orbits (coinciding with the Schwarzschild one for $r\geq 6M$), and is given for comparison. Vertical lines in the corresponding styles indicate the radial positions of $\Omega_{\rm  max}$ test-particle orbital values for $q=1.1$,  1.5 and 1.8, respectively. Diamond symbols in corresponding colors indicate the value of $\Omega(r_0)$ in the simulation in each of the cases.
}
\label{omegas2}
\end{figure*}

\section{Numerical setup}\label{stp}

Aiming to obtain a thin disk in the vicinity of the RN naked singularity, modeled numerically on a grid, we initiate our simulations with a viscous accretion disk, initially placed at a large distance from the NkS, i.e., truncated within some radius far outside the zero-gravity sphere.

Here we briefly present our numerical setup with the publicly available code {\sc pluto} \citep{m07}. The setup is as detailed in \cite{cem19}, the only difference is that now we use the pseudo-Newtonian potential,
Eq.~\ref{rnline2b}, instead of the standard Newtonian one. In a 2D axisymmetric setup in spherical coordinates we set a computational domain with the resolution $R\times\theta=(217\times 200)$ grid cells in $[0.6,30]r_{\rm g}\times[0,\pi]$. In the radial direction
we set a logarithmically stretched grid, and  in the co-latitudinal direction we set a uniform grid with two different resolutions, to better resolve the near-equatorial disk region. From the north pole we set 50 grid cells to $67.5^\circ$, then 100 grid cells to $112.5^\circ$, and then again 50 grid cells to the south pole of the computational domain. The initial disk with the aspect ratio (disk height to cylindrical radius ratio) $\epsilon= h/r = 0.065$, surrounded with an initially non-rotating corona in a hydrostatic equilibrium, is set by the analytical solution from \cite{Kita95,KK00}. The thin disk is initially set up at $r>10M$. Throughout the simulation the disk is fed orbiting fluid at the outer boundary, at a constant accretion rate. The maximal density in the corona is set to 1\% of the maximal disk density. Similar setup, with a thicker $\epsilon=0.1$ disk and uniform $\theta$ grid, was also used in \citet{cembrun23} and \citet{cemklupart23}. The (anomalous) viscosity coefficient is set to a small value $\alpha=5\times 10^{-3}$, for a relatively small accretion rate in the disk. As in all the cases with such a small $\alpha$, a midplane backflow appears in the simulated disk, as described in \cite{MishraRback23}.

Simulations were performed using the second-order piecewise linear reconstruction, with a Van Leer limiter in density and a \texttt{minmod} limiter in pressure and velocity. The second-order time-stepping (\texttt{RK2}) was used, and an approximate Roe solver \citep[and references therein]{m07}, with a modification in the \texttt{flag\_shock} subroutine: flags were set to switch to more diffusive \texttt{hll} solver if the internal energy became lower than 1\% of the total energy, instead of switching in the presence of shocks. We use the polytropic equation of state with the plasma polytropic index $\gamma=5/3$, and the disk is set with all the viscous heating assumed to be locally radiated away from the disk, as appropriate for a thin disk.

The computational domain extends radially down to a fraction of $r_0$, i.e. to a region never penetrated by the accreting fluid; however, we need to specify the inner boundary conditions because of the numerically necessary background low-density density fluid filling the computational domain. The inner radial boundary conditions were set to
\texttt{reflective}, and the outer were set with the constant inflow of material in the disk, with the logarithmic extrapolation in the density and pressure in the coronal part of the domain. The latitudinal boundaries were set to \texttt{axisymmetry}. Units in the {\sc pluto} code are normalized with respect to density, length and velocity scales. In our non-radiative simulations the density scale is arbitrary, and the latter two are in our case naturally assigned as the gravitational radius $r_{\rm g}$ and the speed of light $c$, respectively.
\begin{figure*}[ht!]
\centering 
\includegraphics[width=\columnwidth]{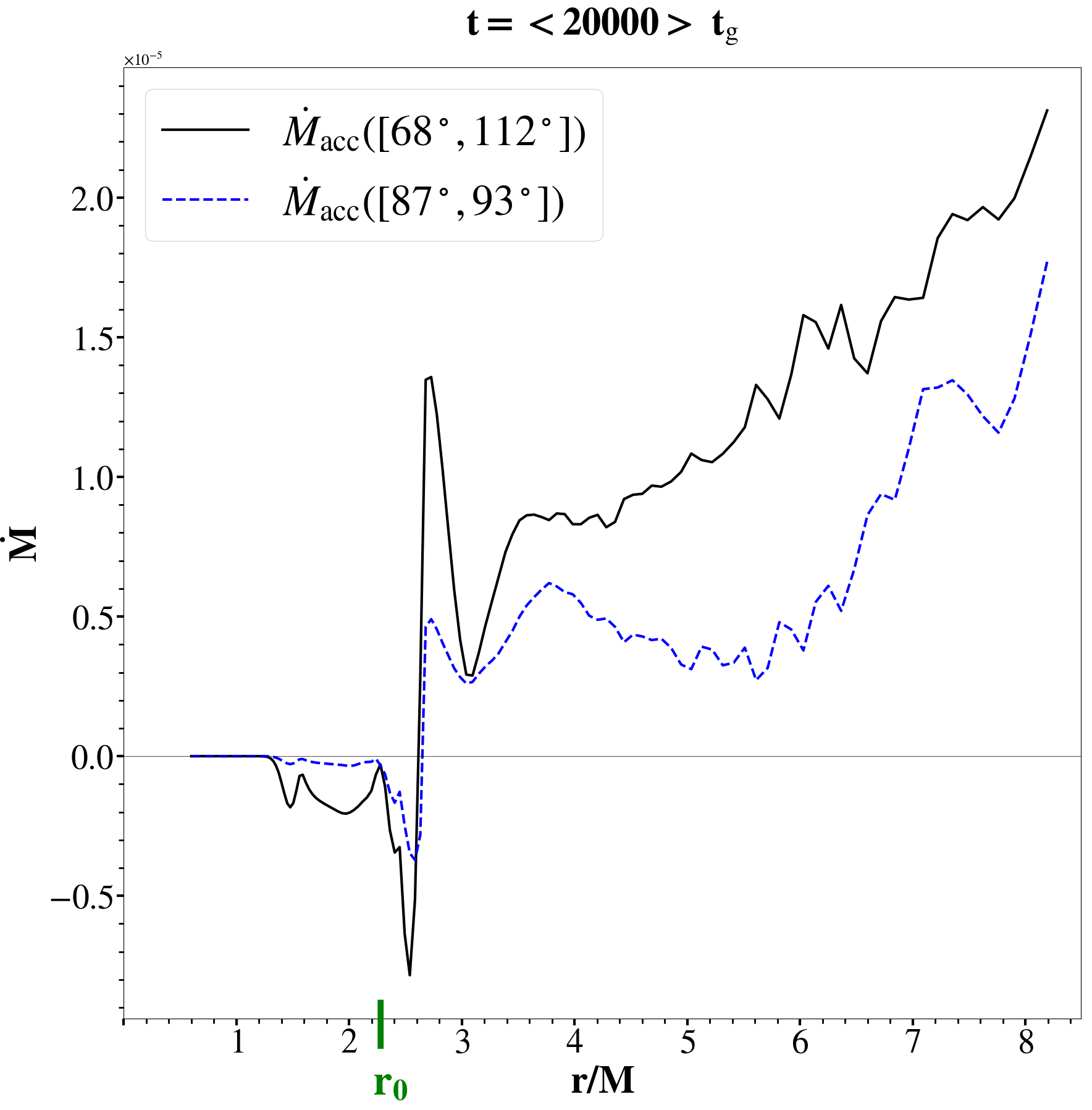}
\includegraphics[width=0.95\columnwidth]{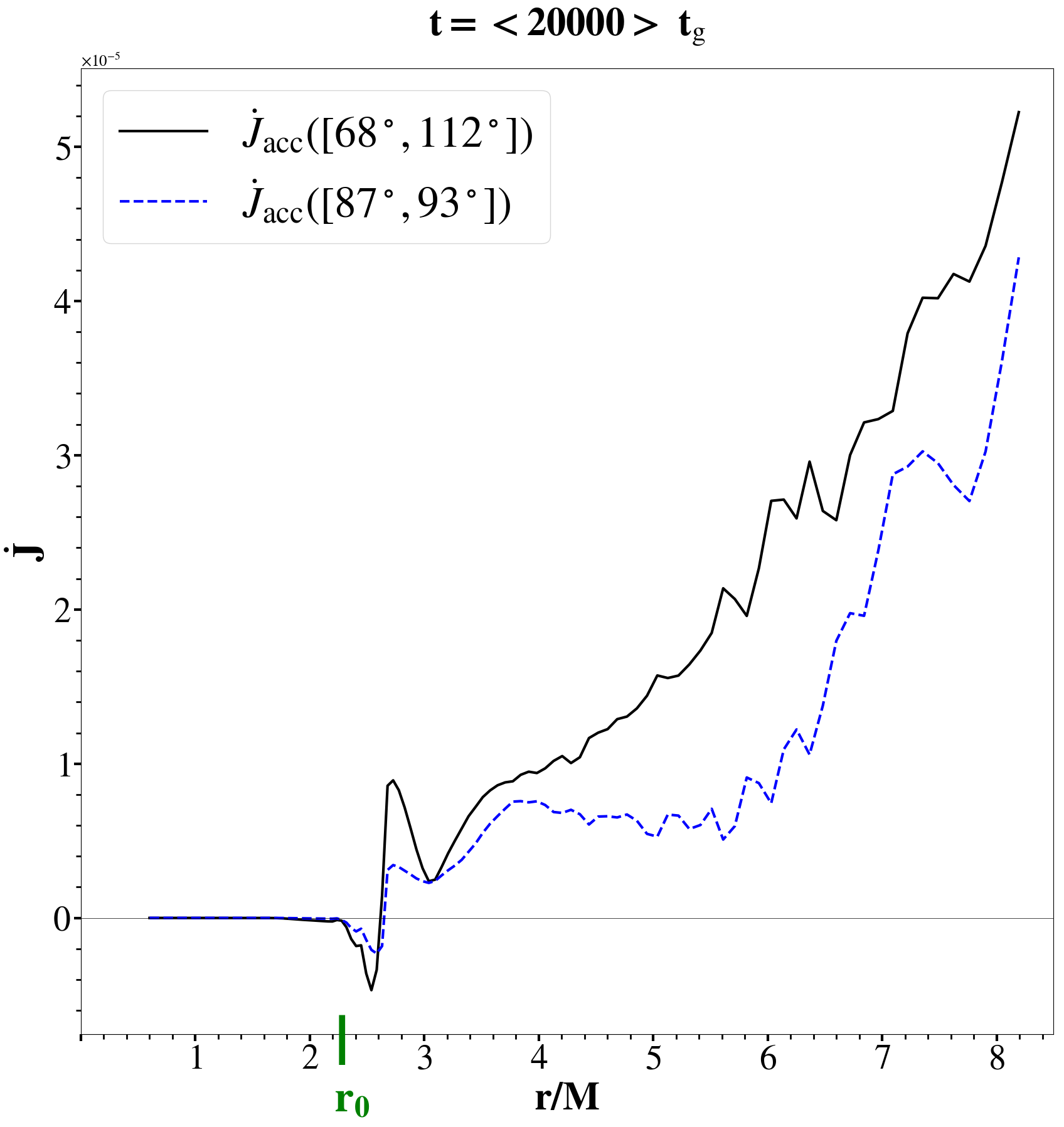}
\caption{Values of $\dot{M}$ and $\dot{J}$ in code units, computed in two co-latitudinal intervals as assigned in the legends, in dependence of radial distance from the origin in the simulation from Fig.~\ref{rho1c5} with $q=1.5$, obtained as an average over the time interval of $t\in [19000, 21000]\ t_{\rm g}$. The  zero-gravity radius $r_0$ is marked with the short green vertical line.}
\label{mjdots1}
\end{figure*}

\section{Results of our simulations}
\label{rezs2}
To investigate the behavior of matter when it approaches the RN naked singularity, we perform numerical simulations in a range of charge parameters $q$ of a RN naked singularity. We focus on the $q=1.5$ case, which exhibits all the characteristics seen also with other values of $q$, as long as $q > \sqrt{9/8}$, i.e., as long as the NkS is in the regime where circular test-particle orbits are stable all the way down to the zero-gravity sphere \citep{MishraKK24}.

The result from an evolved simulation at $t=<20000>~t_{\rm g}$ is shown in Fig.~\ref{rho1c5}. 
The left panel exhibits the density of fluid. One can at once see a fairly thick spherical shell (yellow
in our color scheme) that was formed around the zero gravity sphere ($r_0=2.25M$, here). While this is the non-rotating levitating atmosphere described in \cite{VK23}, in our simulation this structure is not related to the accretion disk at all (at least at the time interval used for computing the average). The shell is a result of radial accretion, both from the outside and the inside of the resulting shell, of the numerically required low-density (non-rotating) background filling the grid; this is immediately apparent from the right panel showing the velocity vectors, as well as from Fig.~\ref{omegas2} which shows the contrast in angular frequency between the rotating disk and the non-rotating background.

We now turn to the thin accretion disk sandwiching the equatorial plane. With a small viscosity, the mass accretion rate is relatively small, and feeding of the disk from the outer boundary is just sufficient to maintain the outer disk in a quasi-steady state (at $r>10M$), while closer to the NkS the mass flow builds up the inner disk which evolves to the structure visible in Figs.~\ref{rho1c5} and \ref{omegas2}. To understand the origin of this structure, we recall that the leading component of the viscous\footnote{We mean here the anomalous effective viscosity, probably of turbulent magnetic origin \citep{SS73,BalbHawl91a}.} stress tensor is proportional to the gradient of the orbital frequency, $T_{r\phi}\propto d\Omega/dr$. 
In a thin accretion disk the orbital frequency is equal to the test-particle value (up to corrections of the order $h/r$).
As long as $d\Omega/dr>0$, as is usual in Newtonian gravity, as well as that of the GR black holes, angular momentum is transported outward by the viscous torques, allowing inward flow of matter (mass accretion). However, for the NkS under discussion here, Keplerian $\Omega(r)$ has a maximum (Fig.~\ref{pl1}). This implies that at radii lower than $\approx (4/3)r_0$ the angular momentum cannot be transported away from a thin disk. Initially, the fluid reaching the location of $\Omega_\mathrm{\max}$ must remain there, while "behind it" (at larger radii) the pressure in the disk builds up. This causes the disk to expand vertically, forming a bulge and the fluid at the inner edge of the disk to approximately adopt the shape of a figure of equilibrium in the effective potential for the appropriate value of the angular momentum  \citep[such figures of equilibrium can be inspected in][]{MishraK23,MishraKK24}. 

Eventually, a sufficient amount of fluid accumulates inside the cylindrical radius of $r\approx(4/3)r_0$ for the angular momentum to be transported by viscous torques from the bulge in the disk to the fluid in that region, allowing accretion in the inner regions of the disk to proceed, with the disk finally extending to the vicinity of the zero-gravity sphere. 
In the left panel of Fiq.~\ref{omegas2} we show a snapshot at $t=20000~t_\mathrm{g}$ of the meridional profile of the angular velocity for the $q=1.5$ simulation. In the right panel we show $\Omega(r)$, of the simulated fluid in the equatorial plane for $q=1.1$, $q=1.5$ and $q=1.8$, as well as (for contrast) the Newtonian expression $\Omega_\mathrm{N}(r)=\sqrt{M/r^3}$.
The equatorial profile in Fiq.~\ref{omegas2} (right panel) of the actual orbital frequency of the fluid, $\Omega(r)$ is a good match to the test-particle orbital frequency $\Omega_\mathrm{RN}(r)$ in Fig.~\ref{pl1} (left panel). To the extent that $\Omega(r)\approx \Omega_\mathrm{RN}(r)$, we can talk of a thin disk approximation even for $r\approx r_0$, although the thickness to radius ratio is approximately $h/r\sim1/2$ at this inner edge.

The mass and angular momentum fluxes $\dot{M}$ and $\dot{J}$ are, respectively:
\begin{equation}
\dot{M}=-2\pi r^2\int_{\theta}\rho \varv_{\rm r} \sin{\theta} d\theta,\quad
\dot{J}=-2\pi r^3\int_{\theta}\rho \varv_{\rm r} \varv_\phi \sin^2{\theta} d\theta .
\label{fluxes}
\end{equation}
The radial dependence for both those fluxes along the meridional plane is shown in Fig.~\ref{mjdots1}. The black solid line shows accretion in a "wedge" of azimuthal angle range $[68^\circ,112^\circ]$, while the blue dashed lines accretion in the azimuthal angle range $[87^\circ,93^\circ]$.  Comparison of the black solid and blue dashed lines in this figure shows that, indeed, accretion through the region of maximum angular frequency occurs at high latitudes, i.e., through the bulge in the disk, as described above.

Thus, our simulation resolves a theoretical difficulty: how does the thin disk negotiate the maximum of the angular momentum, where the torque vanishes? The simulation has found a way: the disk thickens at the zero-torque circle (or cylinder) and the fluid is pushed over the top at higher latitudes (well above and below the equatorial plane). Otherwise, the results with our pseudo-Newtonian potential show a good match to the theoretical predictions: the actual angular velocity has a maximum close to $(4/3)r_0$, and vanishes close to the zero gravity sphere at $r_0$, where it forms a structure reminiscent of rotating figures of equilibrium \citep[c.f.,][]{MishraKK24}.

As expected, the region between the zero gravity radius and the radius of maximum orbital angular velocity is a region of transition between quasi-Newtonian orbital motion and a state of rest near the zero-gravity sphere.
Our simulations in the custom-designed pseudo-Newtonian potential of
Eq.~\ref{rnline2b} seem to capture the expected qualitative features of accretion flow in the RN gravity in this innermost part of the system.

\section{Summary}
\label{conc}

We have performed the first simulations of a thin accretion disk around a naked singularity.
Featuring in various theoretical considerations, naked singularities are vigorously contested as unrealistic by some authors, and appreciated by others as an interesting, and one of the earliest, solution to Einstein's field equations in vacuum. Next to black holes, they describe the most extreme cases of gravitationally collapsed massive objects. Without entering the discussion about their reality, we present numerical simulations of a thin accretion disc around such objects. Accreting compact objects are, of course, the primary sources in X-ray astronomy, and thin accretion disks are ubiquitous. We focus on the Reissner-Nordstr\"om NkS, whose gravity we describe by a custom-designed pseudo-Newtonian potential that reproduces the RN test-particle orbital frequency exactly. We carry out the simulations with the Newtonian hydrodynamics code {\sc pluto}.

We simulate a thin disk with the $\alpha$-viscosity prescription. The main advantage of this is that inclusion of 
the magneto-rotational instability (MRI), would have required a large increase in resolution needed to resolve the MRI, greatly increasing the computational costs. Inclusion of MRI would also require the presence of magnetic fields, which complicate the solutions as well as their interpretation.

We find simulated solutions which compare well with analytic expectations regarding the characteristic shape of the innermost part of the disc near the position of the zero-gravity sphere. Our solutions also preserve the position of the surface where the angular velocity in the disc is maximal. This suggests that our simulations may serve as the foundation for more involved setups, e.g. with the inclusion of the magnetic field, and radiation, at least until such time as our pseudo-Newtonian calculations are superseded by ones in full GR. Our results also open the possibility of studying the stability of a thin disc around naked singularities with different physical parameters in the disc.

\begin{acknowledgments}

Research supported in part by the Polish NCN Grant No. 2019/33/B/ST9/01564. The CHUCK and XL clusters in CAMK Warsaw and ASIAA, Taipei, respectively, are thanked for granting access to Linux computer clusters used for the high-performance computations. M\v{C} acknowledges the Czech Science Foundation (GA\v{C}R) grant No.~21-06825X and the support by the International Space Science Institute (ISSI) in Bern, which hosted the International Team project \#495 (Feeding the spinning top) with its inspiring discussions. The authors thank T. Krajewski for helpful discussions.
\end{acknowledgments}

\bibliography{refspseudo}{}

\begin{thebibliography}{}
\expandafter\ifx\csname natexlab\endcsname\relax\def\natexlab#1{#1}\fi
\providecommand{\url}[1]{\href{#1}{#1}}
\providecommand{\dodoi}[1]{doi:~\href{http://doi.org/#1}{\nolinkurl{#1}}}
\providecommand{\doeprint}[1]{\href{http://ascl.net/#1}{\nolinkurl{http://ascl.net/#1}}}
\providecommand{\doarXiv}[1]{\href{https://arxiv.org/abs/#1}{\nolinkurl{https://arxiv.org/abs/#1}}}

\bibitem[{{Abramowicz}(2009)}]{Abramowicz09}
{Abramowicz}, M.~A. 2009, \aap, 500, 213, \dodoi{10.1051/0004-6361/200912155}

\bibitem[{{Balbus} \& {Hawley}(1991)}]{BalbHawl91a}
{Balbus}, S.~A., \& {Hawley}, J.~F. 1991, \apj, 376, 214,
  \dodoi{10.1086/170270}

\bibitem[{{Buchdahl}(1970)}]{Buch70}
{Buchdahl}, H.~A. 1970, \mnras, 150, 1, \dodoi{10.1093/mnras/150.1.1}

\bibitem[{{Dihingia} {et~al.}(2024){Dihingia}, {Uniyal}, \&
  {Mizuno}}]{Dihingia2024}
{Dihingia}, I.~K., {Uniyal}, A., \& {Mizuno}, Y. 2024, arXiv e-prints,
  arXiv:2410.13406, \dodoi{10.48550/arXiv.2410.13406}

\bibitem[{{EHTC} {et~al.}(2019{\natexlab{a}}){EHTC}, {Alberdi}, {Alef},
  {Asada}, {Azulay}, {Baczko}, {Ball}, {Balokovi{\'c}}, {Barrett}, {Bintley},
  {Blackburn}, {Boland}, {Bouman}, {Bower}, {Bremer}, {Brinkerink},
  {Brissenden}, {Britzen}, {Broderick}, {Broguiere}, {Bronzwaer}, {Byun},
  {Carlstrom}, {Chael}, {Chan}, \& {Chatterjee}}]{M87P1}
{EHTC}, {Akiyama}, K., {Alberdi}, A., {Alef}, W., {et~al.} 2019{\natexlab{a}},
  \apjl, 875, L1, \dodoi{10.3847/2041-8213/ab0ec7}

\bibitem[{{EHTC} {et~al.}(2019{\natexlab{b}}){EHTC}, {Alberdi}, {Alef},
  {Asada}, {Azulay}, {Baczko}, {Ball}, {Balokovi{\'c}}, {Barrett}, {Bintley},
  {Blackburn}, {Boland}, {Bouman}, {Bower}, {Bremer}, {Brinkerink},
  {Brissenden}, {Britzen}, {Broderick}, {Broguiere}, {Bronzwaer}, {Byun},
  {Carlstrom}, {Chael}, {Chan}, \& {Chatterjee}}]{M87P6}
---. 2019{\natexlab{b}}, \apjl, 875, L6, \dodoi{10.3847/2041-8213/ab1141}

\bibitem[{{EHTC} {et~al.}(2019{\natexlab{c}}){EHTC}, {Alberdi}, {Alef},
  {Asada}, {Azulay}, {Baczko}, {Ball}, {Balokovi{\'c}}, {Barrett}, {Bintley},
  {Blackburn}, {Boland}, {Bouman}, {Bower}, {Bremer}, {Brinkerink},
  {Brissenden}, {Britzen}, {Broderick}, {Broguiere}, {Bronzwaer}, {Byun},
  {Carlstrom}, {Chael}, {Chan}, \& {Chatterjee}}]{M87P5}
---. 2019{\natexlab{c}}, \apjl, 875, L5, \dodoi{10.3847/2041-8213/ab0f43}

\bibitem[{{EHTC} {et~al.}(2022{\natexlab{a}}){EHTC}, {Alberdi}, {Alef},
  {Algaba}, {Anantua}, {Asada}, {Azulay}, {Bach}, {Baczko}, {Ball},
  {Balokovi{\'c}}, {Barrett}, {Baub{\"o}ck}, \& {Benson}}]{SgraP1}
---. 2022{\natexlab{a}}, \apjl, 930, L12, \dodoi{10.3847/2041-8213/ac6674}

\bibitem[{{EHTC} {et~al.}(2022{\natexlab{b}}){EHTC}, {Alberdi}, {Alef},
  {Algaba}, {Anantua}, {Asada}, {Azulay}, {Bach}, {Baczko}, {Ball},
  {Balokovi{\'c}}, {Barrett}, {Baub{\"o}ck}, \& {Benson}}]{SgrAP6}
---. 2022{\natexlab{b}}, \apjl, 930, L17, \dodoi{10.3847/2041-8213/ac6756}

\bibitem[{{EHTC} {et~al.}(2022{\natexlab{c}}){EHTC}, {Alberdi}, {Alef},
  {Algaba}, {Anantua}, {Asada}, {Azulay}, {Bach}, {Baczko}, {Ball},
  {Balokovi{\'c}}, {Barrett}, {Baub{\"o}ck}, \& {Benson}}]{SgrAP5}
---. 2022{\natexlab{c}}, \apjl, 930, L16, \dodoi{10.3847/2041-8213/ac6672}

\bibitem[{{GRAVITY Collaboration} {et~al.}(2022){GRAVITY Collaboration},
  {Abuter}, {Aimar}, {Amorim}, {Ball}, {Baub{\"o}ck}, {Berger}, {Bonnet},
  {Bourdarot}, {Brandner}, {Cardoso}, {Cl{\'e}net}, {Dallilar}, {Davies}, {de
  Zeeuw}, {Dexter}, {Drescher}, {Eisenhauer}, {F{\"o}rster Schreiber},
  {Foschi}, {Garcia}, {Gao}, {Gendron}, {Genzel}, {Gillessen}, {Habibi},
  {Haubois}, {Hei{\ss}el}, {Henning}, {Hippler}, {Horrobin}, {Jochum}, {Jocou},
  {Kaufer}, {Kervella}, {Lacour}, {Lapeyr{\`e}re}, {Le Bouquin}, {L{\'e}na},
  {Lutz}, {Ott}, {Paumard}, {Perraut}, {Perrin}, {Pfuhl}, {Rabien},
  {Shangguan}, {Shimizu}, {Scheithauer}, {Stadler}, {Stephens}, {Straub},
  {Straubmeier}, {Sturm}, {Tacconi}, {Tristram}, {Vincent}, {von Fellenberg},
  {Widmann}, {Wieprecht}, {Wiezorrek}, {Woillez}, {Yazici}, \&
  {Young}}]{Gravity2022}
{GRAVITY Collaboration}, {Abuter}, R., {Aimar}, N., {et~al.} 2022, \aap, 657,
  L12, \dodoi{10.1051/0004-6361/202142465}

\bibitem[{{Gyulchev} {et~al.}(2020){Gyulchev}, {Kunz}, {Nedkova}, {Vetsov}, \&
  {Yazadjiev}}]{Gyulchev2020}
{Gyulchev}, G., {Kunz}, J., {Nedkova}, P., {Vetsov}, T., \& {Yazadjiev}, S.
  2020, European Physical Journal C, 80, 1017,
  \dodoi{10.1140/epjc/s10052-020-08575-7}

\bibitem[{{Joshi} {et~al.}(2014){Joshi}, {Malafarina}, \&
  {Narayan}}]{Joshi2014}
{Joshi}, P.~S., {Malafarina}, D., \& {Narayan}, R. 2014, Classical and Quantum
  Gravity, 31, 015002, \dodoi{10.1088/0264-9381/31/1/015002}

\bibitem[{{Kita}(1995)}]{Kita95}
{Kita}, D.~B. 1995, PhD thesis, The University of Wisconsin -- Madison.

\bibitem[{{Kluzniak} \& {Kita}(2000)}]{KK00}
{Kluzniak}, W., \& {Kita}, D. 2000, arXiv e-prints, astro-ph/0006266

\bibitem[{{Klu{\'z}niak} \& {Krajewski}(2024)}]{KK24}
{Klu{\'z}niak}, W., \& {Krajewski}, T. 2024, \prl, 133, 241401,
  \dodoi{10.1103/PhysRevLett.133.241401}

\bibitem[{{MacFadyen} \& {Woosley}(1999)}]{MacFad99}
{MacFadyen}, A.~I., \& {Woosley}, S.~E. 1999, \apj, 524, 262,
  \dodoi{10.1086/307790}

\bibitem[{{Mignone} {et~al.}(2007){Mignone}, {Bodo}, {Massaglia}, {Matsakos},
  {Tesileanu}, {Zanni}, \& {Ferrari}}]{m07}
{Mignone}, A., {Bodo}, G., {Massaglia}, S., {et~al.} 2007, \apjs, 170, 228,
  \dodoi{10.1086/513316}

\bibitem[{{Mishra} \& {Klu{\'z}niak}(2023)}]{MishraK23}
{Mishra}, R., \& {Klu{\'z}niak}, W. 2023, in Proceedings of RAGtime 23-25:
  Workshops on Black Holes and Neutron Stars, ed. Z.~{Stuchlik}, G.~{Torok},
  V.~{Karas}, \& D.~{Lancova}, 151--166

\bibitem[{{Mishra} {et~al.}(2024{\natexlab{a}}){Mishra}, {Krajewski}, \&
  {Klu{\'z}niak}}]{MishraKK24}
{Mishra}, R., {Krajewski}, T., \& {Klu{\'z}niak}, W. 2024{\natexlab{a}}, \prd,
  110, 124030, \dodoi{10.1103/PhysRevD.110.124030}

\bibitem[{{Mishra} {et~al.}(2023){Mishra}, {{\v{C}}emelji{\'c}}, \&
  {Klu{\'z}niak}}]{MishraRback23}
{Mishra}, R., {{\v{C}}emelji{\'c}}, M., \& {Klu{\'z}niak}, W. 2023, \mnras,
  523, 4708, \dodoi{10.1093/mnras/stad1691}

\bibitem[{{Mishra} {et~al.}(2024{\natexlab{b}}){Mishra}, {Vieira}, \&
  {Klu{\'z}niak}}]{MishraRNakSing24}
{Mishra}, R., {Vieira}, R. S.~S., \& {Klu{\'z}niak}, W. 2024{\natexlab{b}},
  \mnras, 530, 3038, \dodoi{10.1093/mnras/stae941}

\bibitem[{{Nojiri} {et~al.}(2017){Nojiri}, {Odintsov}, \&
  {Oikonomou}}]{Nojiri17}
{Nojiri}, S., {Odintsov}, S.~D., \& {Oikonomou}, V.~K. 2017, \physrep, 692, 1,
  \dodoi{10.1016/j.physrep.2017.06.001}

\bibitem[{{Nordstr{\"o}m}(1918)}]{Nords18}
{Nordstr{\"o}m}, G. 1918, Koninklijke Nederlandse Akademie van Wetenschappen
  Proceedings Series B Physical Sciences, 20, 1238

\bibitem[{{Oppenheimer} \& {Snyder}(1939)}]{Openh39}
{Oppenheimer}, J.~R., \& {Snyder}, H. 1939, Physical Review, 56, 455,
  \dodoi{10.1103/PhysRev.56.455}

\bibitem[{{Paczy{\'n}ski} \& {Wiita}(1980)}]{PW80}
{Paczy{\'n}ski}, B., \& {Wiita}, P.~J. 1980, \aap, 88, 23

\bibitem[{{Paugnat} {et~al.}(2022){Paugnat}, {Lupsasca}, {Vincent}, \&
  {Wielgus}}]{Paugnat2022}
{Paugnat}, H., {Lupsasca}, A., {Vincent}, F.~H., \& {Wielgus}, M. 2022, \aap,
  668, A11, \dodoi{10.1051/0004-6361/202244216}

\bibitem[{{Penrose}(1969)}]{Penrose69}
{Penrose}, R. 1969, Riv. Nuovo Cim., 1, 252, \dodoi{10.1023/A:1016578408204}

\bibitem[{{Pugliese} {et~al.}(2011){Pugliese}, {Quevedo}, \&
  {Ruffini}}]{Pugliese11}
{Pugliese}, D., {Quevedo}, H., \& {Ruffini}, R. 2011, \prd, 83, 024021,
  \dodoi{10.1103/PhysRevD.83.024021}

\bibitem[{{Reissner}(1916)}]{Reiss16}
{Reissner}, H. 1916, Annalen der Physik, 355, 106,
  \dodoi{10.1002/andp.19163550905}

\bibitem[{{Shakura} \& {Sunyaev}(1973)}]{SS73}
{Shakura}, N.~I., \& {Sunyaev}, R.~A. 1973, \aap, 24, 337

\bibitem[{{Shapiro} \& {Teukolsky}(1991)}]{Shapiro1991}
{Shapiro}, S.~L., \& {Teukolsky}, S.~A. 1991, \prl, 66, 994,
  \dodoi{10.1103/PhysRevLett.66.994}

\bibitem[{{Starobinsky}(1980)}]{Starob80}
{Starobinsky}, A.~A. 1980, Physics Letters B, 91, 99,
  \dodoi{10.1016/0370-2693(80)90670-X}

\bibitem[{{Stuchl{\'\i}k} {et~al.}(2015){Stuchl{\'\i}k}, {Pugliese}, {Schee},
  \& {Ku{\v{c}}{\'a}kov{\'a}}}]{2015RN}
{Stuchl{\'\i}k}, Z., {Pugliese}, D., {Schee}, J., \& {Ku{\v{c}}{\'a}kov{\'a}},
  H. 2015, European Physical Journal C, 75, 451,
  \dodoi{10.1140/epjc/s10052-015-3663-7}

\bibitem[{{Vagnozzi} {et~al.}(2023){Vagnozzi}, {Roy}, {Tsai}, {Visinelli},
  {Afrin}, {Allahyari}, {Bambhaniya}, {Dey}, {Ghosh}, {Joshi}, {Jusufi},
  {Khodadi}, {Walia}, {{\"O}vg{\"u}n}, \& {Bambi}}]{Vagnozzi23}
{Vagnozzi}, S., {Roy}, R., {Tsai}, Y.-D., {et~al.} 2023, Classical and Quantum
  Gravity, 40, 165007, \dodoi{10.1088/1361-6382/acd97b}

\bibitem[{{{\v{C}}emelji{\'c}}(2019)}]{cem19}
{{\v{C}}emelji{\'c}}, M. 2019, \aap, 624, A31,
  \dodoi{10.1051/0004-6361/201834580}

\bibitem[{{{\v{C}}emelji{\'c}} \& {Brun}(2023)}]{cembrun23}
{{\v{C}}emelji{\'c}}, M., \& {Brun}, A.~S. 2023, \aap, 679, A16,
  \dodoi{10.1051/0004-6361/202243517}

\bibitem[{{{\v{C}}emelji{\'c}} {et~al.}(2023){{\v{C}}emelji{\'c}},
  {Klu{\'z}niak}, \& {Parthasarathy}}]{cemklupart23}
{{\v{C}}emelji{\'c}}, M., {Klu{\'z}niak}, W., \& {Parthasarathy}, V. 2023,
  \aap, 678, A57, \dodoi{10.1051/0004-6361/202140637}

\bibitem[{{Vieira} \& {Klu{\'z}niak}(2023)}]{VK23}
{Vieira}, R. S.~S., \& {Klu{\'z}niak}, W. 2023, \mnras, 523, 4615,
  \dodoi{10.1093/mnras/stad1718}

\bibitem[{{Vieira} {et~al.}(2014){Vieira}, {Schee}, {Klu{\'z}niak},
  {Stuchl{\'\i}k}, \& {Abramowicz}}]{VSK14}
{Vieira}, R. S.~S., {Schee}, J., {Klu{\'z}niak}, W., {Stuchl{\'\i}k}, Z., \&
  {Abramowicz}, M. 2014, \prd, 90, 024035, \dodoi{10.1103/PhysRevD.90.024035}

\bibitem[{{Vincent} {et~al.}(2021){Vincent}, {Wielgus}, {Abramowicz},
  {Gourgoulhon}, {Lasota}, {Paumard}, \& {Perrin}}]{Vincent2021}
{Vincent}, F.~H., {Wielgus}, M., {Abramowicz}, M.~A., {et~al.} 2021, \aap, 646,
  A37, \dodoi{10.1051/0004-6361/202037787}

\bibitem[{{Virbhadra} \& {Ellis}(2002)}]{VirbhadraEllis2002}
{Virbhadra}, K.~S., \& {Ellis}, G.~F. 2002, \prd, 65, 103004,
  \dodoi{10.1103/PhysRevD.65.103004}

\bibitem[{{Virbhadra} {et~al.}(1998){Virbhadra}, {Narasimha}, \&
  {Chitre}}]{Virbhadra1998}
{Virbhadra}, K.~S., {Narasimha}, D., \& {Chitre}, S.~M. 1998, \aap, 337, 1,
  \dodoi{10.48550/arXiv.astro-ph/9801174}

\bibitem[{{Wagner}(2023)}]{Wagner2023}
{Wagner}, J. 2023, arXiv e-prints, arXiv:2301.13210,
  \dodoi{10.48550/arXiv.2301.13210}

\bibitem[{{Wagner}(2024)}]{Wagner2024}
---. 2024, arXiv e-prints, arXiv:2405.08057, \dodoi{10.48550/arXiv.2405.08057}

\bibitem[{{Wielgus}(2021)}]{Wielgus2021}
{Wielgus}, M. 2021, \prd, 104, 124058, \dodoi{10.1103/PhysRevD.104.124058}

\bibitem[{{Woosley}(1993)}]{Woosley93}
{Woosley}, S.~E. 1993, \apj, 405, 273, \dodoi{10.1086/172359}

\end{thebibliography}
\bibliographystyle{aasjournal}

\end{document}